\documentclass[aps,prl,amsmath,amssymb,showpacs,reprint,floatfix,superscriptaddress]{revtex4-1}

\usepackage{graphicx}
\usepackage{color}
\usepackage[normalem]{ulem}
\usepackage{cancel}

\newcommand{\Ein}{\vec{E}_{{\textrm{in}}}}
\newcommand{\Escat}{\vec{E}_{{\textrm{scat}}}}

\allowdisplaybreaks[2]

\begin{document}

\author{K. P. Heeg}
\affiliation{Max-Planck-Institut f\"{u}r Kernphysik, Saupfercheckweg 1, D-69117 Heidelberg, Germany}

\author{C. Ott}
\affiliation{Max-Planck-Institut f\"{u}r Kernphysik, Saupfercheckweg 1, D-69117 Heidelberg, Germany}

\author{D. Schumacher}
\affiliation{Deutsches Elektronen-Synchrotron DESY, Notkestra{ss}e 85, 22607 Hamburg, Germany}

\author{H.-C. Wille}
\affiliation{Deutsches Elektronen-Synchrotron DESY, Notkestra{ss}e 85, 22607 Hamburg, Germany}

\author{R. R\"ohlsberger}
\affiliation{Deutsches Elektronen-Synchrotron DESY, Notkestra{ss}e 85, 22607 Hamburg, Germany}

\author{T. Pfeifer}
\affiliation{Max-Planck-Institut f\"{u}r Kernphysik, Saupfercheckweg 1, D-69117 Heidelberg, Germany}

\author{J. Evers}
\affiliation{Max-Planck-Institut f\"{u}r Kernphysik, Saupfercheckweg 1, D-69117 Heidelberg, Germany}

\title{Interferometric phase detection at x-ray energies via Fano resonance control}

\begin{abstract}
Modern x-ray light sources promise access to structure and dynamics of matter in largely unexplored spectral regions. However, the desired information is encoded in the light intensity and phase, whereas detectors register only the intensity. This phase problem is ubiquitous in crystallography and imaging, and impedes the exploration of quantum effects at x-ray energies.
Here, we demonstrate phase-sensitive measurements characterizing the quantum state of a nuclear two-level system at hard x-ray energies. The nuclei are initially prepared in a superposition state. Subsequently, the relative phase of this superposition is interferometrically reconstructed from the emitted x-rays.
Our results form a first step towards x-ray quantum state tomography, and provide new avenues for structure determination and precision metrology via x-ray Fano interference.
\end{abstract}

\date{\today}


\maketitle

The phase of electromagnetic fields is the key to interferometry. This technique of superimposing electromagnetic waves is an important method with applications across all the natural sciences~\cite{interferometer}. But standard detectors for optical or higher-frequency fields are sensitive to the field intensity only, masking the phase required for many applications~\cite{zavetta}. In x-ray science, prominent examples are crystallography~\cite{nanocrystal} and coherent imaging. In the latter, photons are elastically scattered off of an object to characterize it~\cite{image-review}. Although the relation between the scattered light and the original object is well known, the lack of phase information prevents a straightforward reconstruction of the original object~\cite{Taylor:ba5147}. 
Next to structure determination, also the reconstruction of quantum states requires phase-sensitive measurements~\cite{tomography,PhysRevA.40.2847}. This quantum state tomography has been successfully demonstrated at optical frequencies~\cite{PhysRevLett.70.1244}, but remains an open challenge for the emerging field of x-ray quantum optics~\cite{xray,cls,RR12,sgc,olga,LASER,MIXING,STIM,Heeg2014}. 

In this work, we demonstrate phase-sensitive measurements on an archetype quantum mechanical two-level system (TLS) at hard x-ray frequencies, represented by a nuclear resonance of a M\"ossbauer isotope. The TLS is realized in a large ensemble of identical nuclei, operated in such a way that the incident x-rays couple the ground state to a single collective excited state.
The phase-sensitivity is gained by a cavity-based x-ray interferometer with the TLS in one of its arms. Tuning the phase of the non-TLS path enables us to determine the phase of the light emitted by the TLS via the intensity at one of the output ports of the interferometer. 
The x-ray pulse is near-instantaneous on the time scale of the nuclei, and prepares the TLS in an initial coherent superposition of its two states. After this,
we determine the phase of the light
emitted by the TLS, which allows us to characterize the initial state prepared by the x-ray pulse.

The state of the TLS is described by its density matrix, and the phase reconstructed here can be identified with the phase of the off-diagonal density matrix elements. Our measurement therefore forms an important step towards the full tomography of the TLS quantum state, which is a crucial tool for the exploration of quantum effects at x-ray energies. The interferometric measurement technique combined with the precise mapping of the spectroscopic lineshape of the TLS in turn provide
new avenues for structure determination and precision metrology.

%
%
\begin{figure*}[t]
	\centering
	\includegraphics[width=16cm]{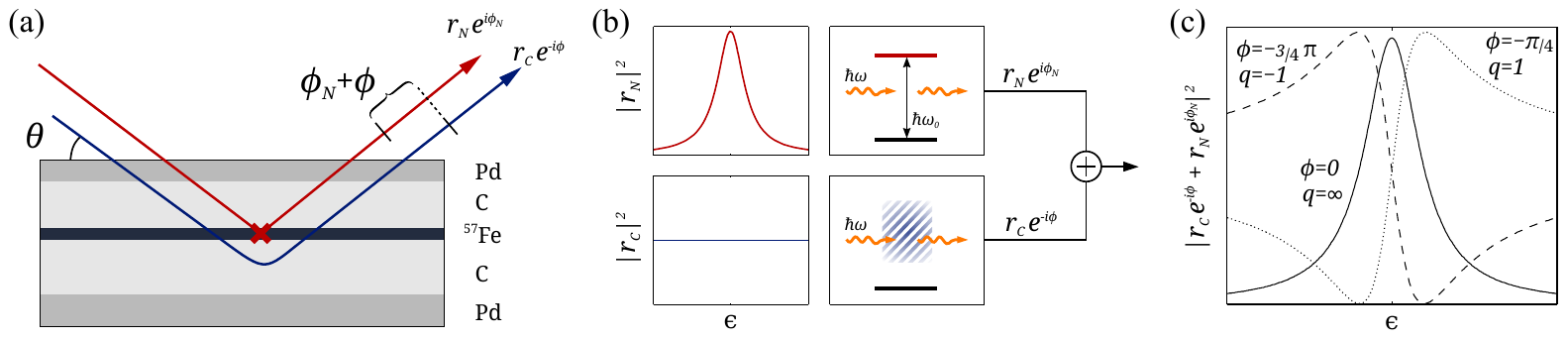}
\caption{(Color online) Schematic setup of the x-ray interferometer and origin of the Fano interference. (a) X-ray light is reflected by a thin-film cavity under a grazing angle $\theta$. The Pd layers act as cavity mirrors and C as the guiding layer. The phases of the reflection by the empty cavity ($r_C e^{-i\phi}$, blue) and from the embedded $^{57}$Fe nuclei in the center ($r_N e^{i\phi_N}$, red) can be controlled via the incidence angle $\theta$ and the energy $\epsilon$, respectively. (b) The empty-cavity reflection forms a broad spectral continuum, whereas the isolated ``bound state'' nuclear response $r_N$ has a Lorentzian shape. (c) The interference of these two paths leads to asymmetric line shapes controlled by the phase $\phi$, which we observe experimentally.}
	\label{fig:setup}
\end{figure*}

The experimental scheme and setup are illustrated in Fig.~\ref{fig:setup}. A nm-sized thin-film structure of materials with alternating index of refraction is used to form an x-ray cavity. The guiding layer of the cavity contains a sheet of ${}^{57}$Fe nuclei with a M\"ossbauer resonance at $14.4$~keV, which is used to form the TLS. X-ray light impinges on this structure in grazing incidence at angle $\theta$, and we record the spectrum of the reflected light. 
In this setting, we interpret the cavity as an interferometer, comprising two pathways that contribute to the x-ray reflectance as depicted in Fig.~\ref{fig:setup}. The first path ($r_C e^{-i\phi}$) consists of light reflected by the cavity alone. In the second path ($r_N e^{i\phi_N}$), the x-rays interact with the near-resonant TLS. The total recorded reflectance arises from the interference between the two paths, governed by their relative phase. As we will show, the phase $\phi$ depends on the incidence angle $\theta$, and thus can externally be controlled. 

We have experimentally explored this phase control at the Dynamics Beamline P01 of the PETRA III synchrotron radiation source (DESY, Hamburg). 
We employed nuclear resonant scattering, where a short broadband incident pulse excites the nuclei, and subsequently the delayed scattered photons are detected in a time window $40-190$ ns after excitation. The cavity is formed by a Pd(4 nm)/C(36 nm)/Pd(14 nm) (top to bottom) layer system with the Pd layers acting as the mirrors and the C as guiding layer. A 1.2 nm thick active layer of ${}^{57}$Fe was placed in the center of the carbon layer. Far-off-resonant background photons are suppressed using a high-resolution monochromator for the incident light. To record the spectrum of the scattered light, we use a spectrally narrow resonant absorber foil (consisting of a 6 $\mu$m thick stainless steel foil enriched to $95\%$ in $^{57}$Fe), which we scan in energy across the nuclear resonance with the help of a Doppler drive~\cite{cls}. The M\"ossbauer nucleus ${}^{57}$Fe features a transition at $\hbar\omega_0=14.4$ keV with a line width $\hbar\gamma = 4.7$ neV. 
Since the incident x-ray pulse in the 10-100~ps range is much shorter than the natural nuclear life time of $141$~ns, and since we record the light emitted from the nuclei starting a few ten nanoseconds after excitation by the x-ray pulse, the preparation of the TLS state in the low-excitation regime and the measurement of the subsequently emitted light can be approximated as independent processes.

\begin{figure}[t!]
	\centering
	\includegraphics[width=8cm]{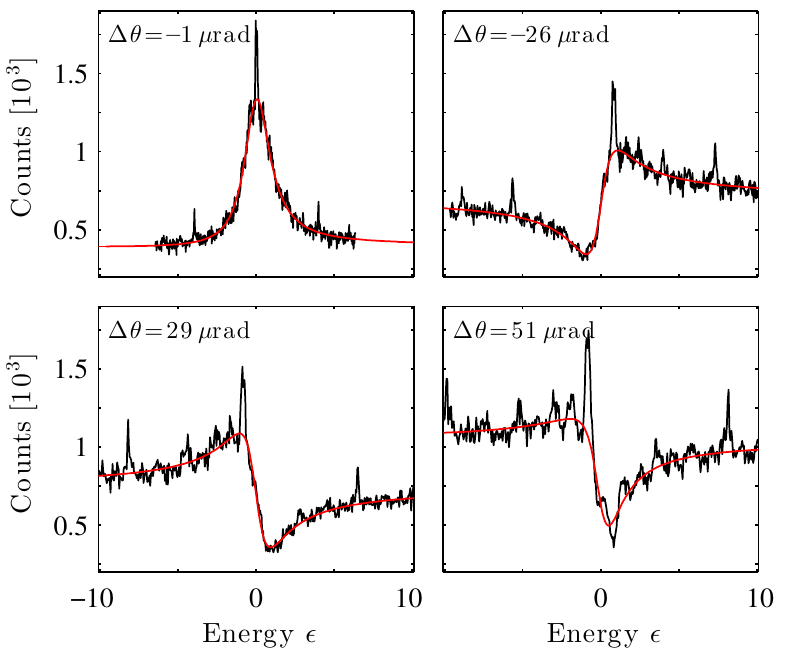}
	\caption{(Color online) Fano line shape control with nuclei. The different panels show the reflected intensity measured at different relative incidence angles $\Delta \theta = \theta - \theta_\textrm{min}$. 
	Experimental raw data is shown without baseline subtraction. Therefore the intensities in the different panels cannot directly be compared.
	}
	\label{fig:res1}
\end{figure}

Measured emission spectra are shown in Fig.~\ref{fig:res1} for selected incidence angles $\theta$ around $\theta = \theta_\textrm{min}$, where the cavity reflectance assumes a deep minimum at frequencies far off the nuclear resonance. Spectra covering a wider range of angles are summarized in the Supplemental Material~\cite{supplement}.
Clearly, the incidence angle acts as a knob to control the spectral response from a Lorentzian shape for $\theta = \theta_\textrm{min}$ to strongly asymmetric line shapes, demonstrating the interferometric nature of our setup.

To interpret these results, we follow a recently developed quantum-optical framework for the description of nuclei in x-ray waveguides, and obtain at critical coupling~\cite{qomodel} for the experimentally observed reflectance
\begin{equation}
|R|^2 =   \left | \sqrt{\sigma_0} \: e^{-i\phi} + \: 
\frac{\gamma_\textrm{SR} }{\Gamma} \: \frac{1}{\epsilon + i}
\right |^2 \label{eq:fano}
\,.
\end{equation}
Here, the first term corresponds to the interferometer path due to the cavity alone, with empty-cavity response $\sigma_0 = 1/(1 + \kappa^2/\Delta_C^2)$, with cavity loss rate $\kappa$, and $\Delta_C$ is the detuning between cavity eigenmode frequency and the frequency of the probing x-ray field.
The second path is represented by a complex-valued Lorentzian bound state amplitude typical of a TLS in the second term. The dimensionless energy $\epsilon=(\omega - \omega_0 -\Delta_\textrm{LS})/(\Gamma/2)$ is modified by cooperative phenomena: The Lorentzian is not centered on the nuclear resonance at $\omega_0$, but slightly shifted by $\Delta_\textrm{LS}$ due to a collective Lamb shift~\cite{cls}. Moreover, its width $\Gamma$ is superradiantly broadened from the natural line width $\hbar\gamma=4.7$ neV of $^{57}$Fe to $\Gamma = \gamma + \gamma_\textrm{SR}$. These cooperative modifications to the resonance position and width are given by $\Delta_\textrm{LS} =$ $-(2/3)\,|g|^2\,N\, \Delta_C/(\kappa^2 + \Delta_C^2)$ and $\gamma_\textrm{SR} =$ $(4/3)\,|g|^2\,N\, \kappa/(\kappa^2 + \Delta_C^2)$, respectively~\cite{qomodel}. Here, $g$ is the cavity-nucleus coupling constant, and $N$ the number of nuclei.

The experimentally observed line shapes can then be understood by noting that the TLS response features the narrow spectral width typical of M\"ossbauer resonances, whereas the cavity modes have orders of magnitude higher spectral width and therefore act as continuum channels. The interference of a narrow bound state with a continuum is known to give rise to asymmetric Fano resonances~\cite{fano}. Relating the relative phase $\phi$ and the Fano $q$ parameter as $\phi = \arg (q - i)$~\cite{universal-fano}, we can indeed rewrite Eq.~(\ref{eq:fano}) as a Fano profile
\begin{equation}
\label{eq:fanoq}
|R|^2 =  \frac{|\epsilon + q|^2}{1+\epsilon^2} \: \sigma_0 \,, \quad
q = \frac{\gamma_\textrm{SR}}{\Gamma}\, \frac{\kappa}{\Delta_C} + i \, \frac{\gamma}{\Gamma}\,.
\end{equation}
We find that $q$ has an imaginary component, which corresponds to an incoherent loss channel~\cite{loss2}. However, in the strongly superradiant case $\Gamma \gg \gamma$, and the loss channels can be neglected. Then, $q\approx \kappa/\Delta_C$, and $|R|^2 \approx (\epsilon+q)^2/[(1+\epsilon^2)(1+q^2)]$, such that $0\leq |R|^2\leq 1$. In particular in this limit, the present system forms an ideal implementation of the original Fano model.
In contrast, in the opposite limit $\Gamma \approx \gamma$, we find $|R|^2 \approx \sigma_0$ without any spectral signatures. It is therefore the collectively enhanced decay rate which enables the Fano implementation with full visibility of the reflectance modulation despite the low $Q$ factor of the cavity.
Note that asymmetric line shapes for nuclear resonances at x-ray energies have previously been predicted or observed~\cite{EarlyFano1,EarlyFano2,EarlyFano3,EarlyFano5,EarlyFano6,Shvydko_exitions_bragg}, though not interpreted as Fano resonances.
Close to the cavity resonance at $\Delta_C = 0$, one can linearize $\Delta_C \approx \delta_C  (\theta - \theta_\textrm{min})$, where $\theta_\textrm{min}$ is the incidence angle where the reflectance of the cavity alone vanishes.
Thus, the incidence angle $\theta$ can be used to control $\Delta_C$, and thereby $q$ and the interferometer phase $\phi$.

For a quantitative analysis of the experimental data, we fitted a generic Fano line shape to the experimentally recorded spectra. Each fit was repeated multiple times with randomly modified initial parameters to avoid bias, and the respective results are indistinguishable within their error bars.
This procedure enables us to determine the superradiant decay width $\Gamma$, the cooperative Lamb shift $\Delta_\textrm{LS}$, as well as the Fano $q$ parameter as a function of the incidence angle $\theta$ independent of our theoretical model. As can be seen from Fig~\ref{fig:res1}, we found good quantitative agreement of model and data. 
Using the such obtained superradiant enhancement $\gamma_\textrm{SR}$ and the cooperative Lamb shift $\Delta_\textrm{LS}$, we normalized the experimental spectra to the dimensionless energy $\epsilon$ for further analysis (see Supplemental Material~\cite{supplement} for other recorded spectra).

A full state tomography requires a measurement of the TLS density matrix~\cite{tomography}. Due to a lack of intensity calibration, we can determine the off-diagonal density matrix elements up to a global scaling factor in the present experiment. The x-ray photons are coherently scattered, preserving their energy. Selecting all detection events of a particular photon energy $\epsilon$ therefore provides access to a large number of identically prepared TLS states. Repeated measurements on the light emitted by identically prepared TLS states then enables us to determine the characteristics of the off-diagonal density matrix element $\rho_{eg}$. 
Up to a global scaling factor, 
\begin{equation}
 \rho_{eg} \sim \sigma_{eg}(\epsilon) \cdot e^{i \phi_N} \;,
\end{equation}
where $\sigma_{eg}(\epsilon)$ contains the spectral shape and $\phi_N$ is the phase of the density matrix element. The shape is directly obtained from the pure nuclear spectrum shown in the top left panel of Fig.~\ref{fig:res1}, where the empty-cavity response vanishes at $\theta = \theta_\textrm{min}$. Further, our interferometric measurements provide a handle to determine the desired phase of the off-diagonal density matrix elements, since $\rho_{eg}$ is directly proportional to the light amplitude emitted by the TLS (see Supplemental Material~\cite{supplement}). Hence, the TLS phase can be identified with the phase $\phi_N$ of the off-diagonal density matrix elements.

\begin{figure}[t]
	\centering
	\includegraphics[width=8cm]{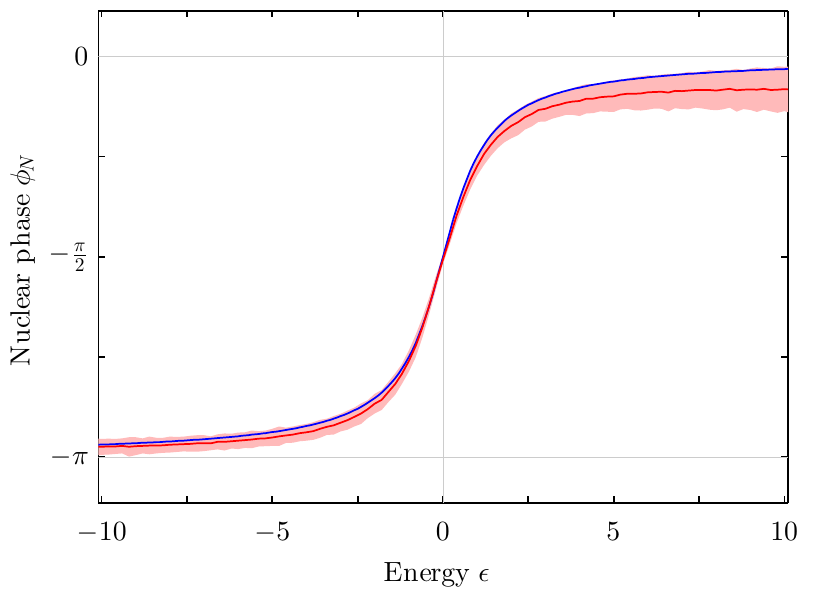}
\caption{(Color online) Nuclear phase reconstructed from the experimental data. The red line shows the phase obtained via Eq.~(\ref{eq:xi}) as a function of the scaled energy $\epsilon$. Error ranges are depicted in light red color. The theoretically expected phase of the Lorentzian typical for a TLS is shown in blue.
}
	\label{fig:phase}
\end{figure}
For the reconstruction of the phase of the nuclear contribution, i.e.~the phase of the TLS, we employed a general ansatz for the reflectance 
\begin{equation}
|R(\Delta\theta, \epsilon)|^2 \sim |r_C(\Delta\theta) e^{-i\phi} + r_N(\epsilon) e^{i\phi_N}|^2\,,
\end{equation}
without assumptions on the shape of the nuclear response $r_N$ except that it only depends on $\epsilon$ and that it vanishes at large detunings $r_N(\epsilon \rightarrow \pm\infty) = 0$. To extract the nuclear phase $\phi_N$, we define the experimentally accessible quantity
\begin{align}
 \xi(\Delta\theta, \epsilon) &= \frac{|R(\Delta\theta, \epsilon)|^2 - |R(0, \epsilon)|^2 - |R(\Delta\theta, \pm\infty)|^2}{2 |R(0, \epsilon)| |R(\Delta\theta, \pm\infty)|} \nonumber \\ 
&= \cos{(\phi + \phi_N)}\,. \label{eq:xi}
\end{align}
As the relation between the incidence angle and the cavity phase $\phi(\Delta\theta) = \arg{[q(\Delta\theta) - i]}$ is known from the quantum optical model, the phase of the nuclear contribution $\phi_N$ can then be determined as function of $\epsilon$ via Eq.~(\ref{eq:xi}) by fitting the cosine to the measured $\xi$ for all available $\Delta\theta$ values.
%
%
To evaluate $\xi(\Delta\theta,\epsilon)$ from the data without referring to the line shape to be reconstructed, we fitted a general rational function
$R_\textrm{rat} = (a_0+a_1 \epsilon + a_2 \epsilon^2)/(b_0+b_1 \epsilon + b_2 \epsilon^2)$
to the data, normalized it to $0 \le R_\textrm{rat} \le 1$, and evaluated it at the according values for $\Delta\theta$ and $\epsilon$. Since the angle $\Delta\theta=0$ was not measured, we obtained $|R(0, \epsilon)|$ from the mean of the results for $\Delta\theta = \pm 1~\mu$rad.

Results of this phase retrieval are shown in Fig.~\ref{fig:phase}. It can be seen that the reconstructed phase of the off-diagonal density matrix element agrees well to the expected Lorentzian shape as function of the dimensionless energy $\epsilon$, in particular close to the resonance energy. 
The main cause of the discrepancy with respect to the expected phase at large $\epsilon$ can be traced back to an uncertainty of $|R(0, \epsilon)|$ in the denominator of Eq.~(\ref{eq:xi}). Here, the measured values are tiny and already small absolute deviations result in large relative errors.

The phase-sensitive interferometric measurement of the optical response of a TLS demonstrated here opens a number of promising research directions. On the one hand, combination of the techniques developed here with measurements of the magnitude of the density matrix elements, either via detecting light intensity or conversion electrons~\cite{PhysRevB.53.171}, could lead to the development of complete quantum state tomography at x-ray energies. Importantly, the method demonstrated here does not depend on the Lorentzian line shape, but can be used to reconstruct the phase of arbitrary nuclear line shapes. Therefore, also more advanced setups, e.g., involving multiple magnetic hyperfine states with selectively coupled resonances can be addressed~\cite{RR12,sgc}. 
On the other hand, the nuclear resonances are of primary significance in precision spectroscopy and metrology at x-ray frequencies due to their narrow line width~\cite{PhysRevLett.3.439}. Using our approach, tiny phase changes can be extracted from the measured data with high precision, assisted by the discovery of the mechanism behind the asymmetric line shapes. 
Conversely, the interferometric phase can be used to manipulate light--matter interactions, as demonstrated by the Fano line shape control which enables us to continuously adjust between Lorentz and Fano line shapes in the x-ray optical response. The high sensitivity of the Fano line shape on the arrangement of scatterers allows for a multitude of applications ranging from structure determination with unprecedented accuracy to precision stabilization of interferometers. Furthermore, Fano interferences are ubiquitous features in light--matter interaction, and our phase control concept provides access to such large application potential of Fano processes in the x-ray region~\cite{nanofano1,nanofano2}. These concepts can also be generalized towards active and dynamical control of spectroscopic line shapes~\cite{universal-fano}, further fueling the emerging field of x-ray quantum optics.

\begin{acknowledgments}
We are grateful to T.~Guryeva for support during sample preparation and to F.~U.~Dill and K.~Schlage for setting up beamline instrumentation. KPH acknowledges funding by the German National Academic Foundation. TP acknowledges support from the MPRG program of the Max-Planck-Gesellschaft (MPG). 
\end{acknowledgments}

\bibliographystyle{myprsty}
\bibliography{nuclear-fano}

\clearpage


\renewcommand{\theequation}{S\arabic{equation}}
\renewcommand{\thefigure}{S\arabic{figure}}
\setcounter{equation}{0}
\setcounter{figure}{0}

{\large{\bf Supplemental Material to: Interferometric phase detection at x-ray energies via Fano resonance control}}

\section{Reflection spectra}
\begin{figure*}[t!]
	\centering
	\includegraphics[width=0.95\textwidth]{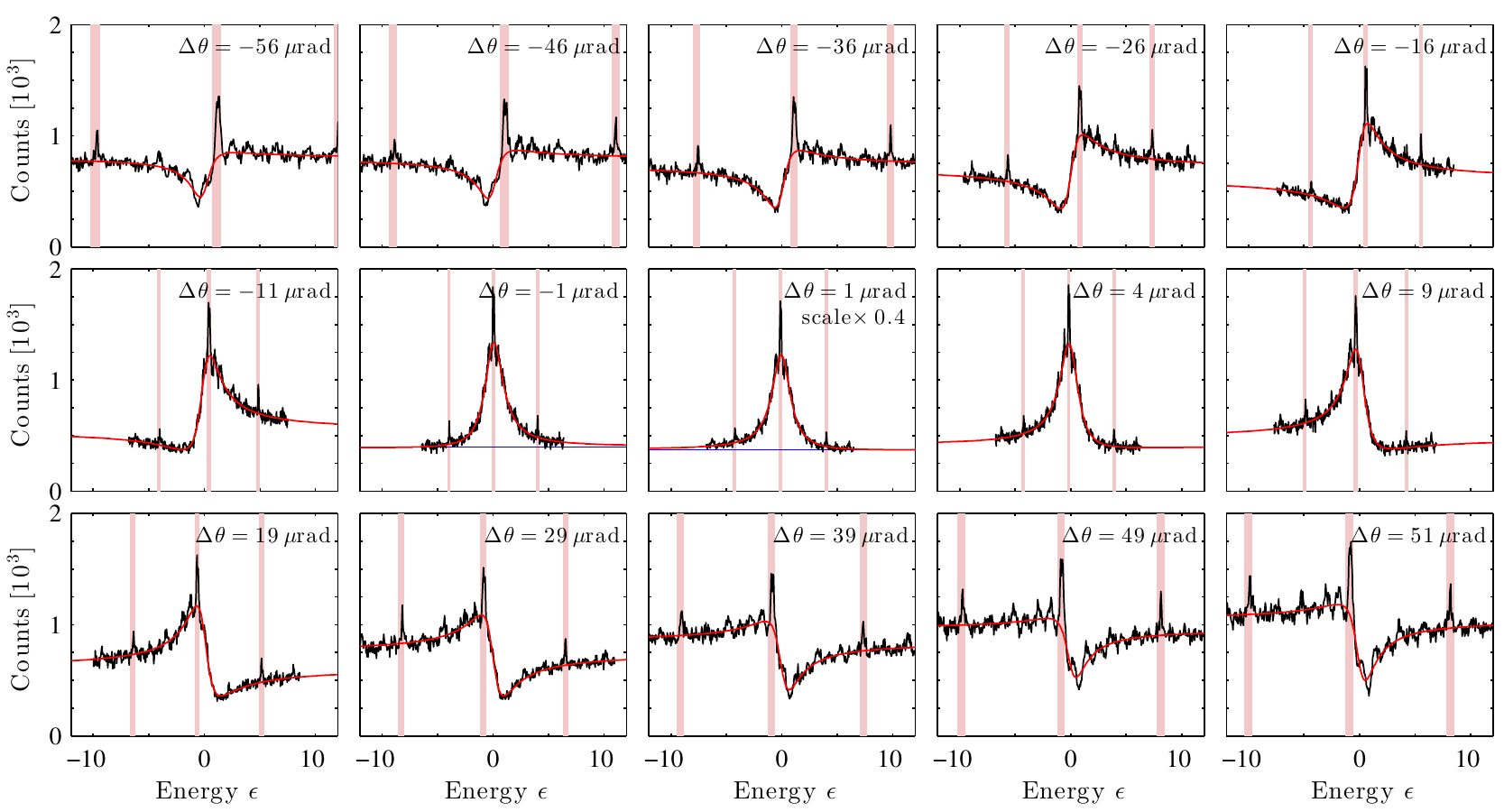}
	\caption{Cavity reflectance for several relative incidence angles $\Delta\theta$ of the probing x-ray field. Black line shows experimental data, overlayed red curves are theory fits. The narrow spikes at the red shaded areas visible in all panels are artefacts of the measurement procedure and have been excluded from the data analysis. The blue horizontal lines at $\Delta\theta = \pm 1 \,\mu$rad are a guide to the eye and indicate the slight asymmetry of the line shapes.}
	\label{fig:all_spectra}
\end{figure*}

Spectra covering a wider range of angles are summarized in Fig.~\ref{fig:all_spectra}. The red shaded areas show spectral regions overlayed with artefacts from the measurement procedure. These regions are excluded from the data analysis. 

\section{Derivation of the cavity reflectance}
For the derivation of Eqs.~(1) and (2) in the main text, we followed a recently developed quantum optical framework for the description of nuclei in x-ray waveguides~\cite{qomodel}. We start with the full model including the large ensemble of nuclei coupled to the quantized cavity field. As the magnetic field experienced by the nuclei is approximately zero, polarization and magnetization effects are insignificant, and we consider a single incident field mode $a_{in}$. The corresponding mode within the cavity is modelled by field operator $a$, the reflected mode observed by the detector by operator $a_{out}$. The reflectance accessible experimentally can then be evaluated as $|R|^2 = |\langle a_{out}\rangle/\langle a_{in}\rangle|^2$.
Making use of the large cavity loss rate $\kappa$ typical in current experiments, we adiabatically eliminate the quantized cavity modes to derive an effective master equation for the nuclei alone. $\kappa$ is a fixed parameter determined by the material composition and the geometric structure of the cavity. Next, motivated by the low resonant intensity of the incident x-ray beam, we evaluate the linear response of the nuclei. By introducing a suitable many-body basis, the nuclear ensemble of $N$ atoms can be modelled as an effective two-level system with parameters modified by cooperative effects. Using standard input-output formalism for cavities~\cite{io}, the output field can be evaluated as $a_{out} = -a_{in} + \sqrt{2\kappa_{R}} \: a$. Here, $\kappa_{R}$ is a coupling parameter quantifying the in- and out-coupling of cavity light. As shown in~\cite{qomodel}, the reflectivity of the single unmagnetized iron layer then follows as $R = R_C + R_N$ with
\begin{eqnarray}
R_C &=&  -1 + \frac{2\kappa_{R}}{\kappa + i\Delta_C} \;, \label{supp_eqn:R_C}\\
R_N &=& - \frac{2i\kappa_{R}}{(\kappa + i\Delta_C)^2} \frac{\frac 23 |g|^2N }{(\Delta-\Delta_\textrm{LS}) + \frac{i}{2}\Gamma} \;, \label{supp_eqn:R_N} \\
\Delta_\textrm{LS} &=& -\frac 23\,|g|^2\,N\, \frac{\Delta_C}{\kappa^2 + \Delta_C^2}\;, \\
\Gamma &=& \gamma+\gamma_\textrm{SR}\;, \\
\gamma_\textrm{SR} &=& \frac 43 \,|g|^2\,N\, \frac{\kappa}{\kappa^2 + \Delta_C^2} \;.
\end{eqnarray}
$R_C$ is the electronic reflection from the cavity alone, as it would be observed in the absence of resonant nuclei.
$R_N$ is the contribution of the nuclei, with atom-cavity coupling $g$. The Fano line shapes arise from the interference of these two contributions.

Finally, we specialize to critical coupling $\kappa = 2\kappa_{R}$. In this case, if the empty cavity is driven by the external field on resonance (cavity detuning $\Delta_C = 0)$, then the reflection from the cavity alone is zero, $R_C = 0$. Experimentally, this condition is achieved, e.g., by choosing a suitable thickness of the top layer of the waveguide. Exploiting critical coupling, the cavity reflectance can be rewritten without further approximations as in Eqs.~(1) and (2).

\section{Relation between the phase $\phi$ and the Fano $q$ parameter\label{sec:supp_sec_q_phi}}
As explained in the main text the Fano asymmetry parameter $q$ can be mapped to a relative phase $\phi$ between the continuum channel and the bound state amplitude~\cite{universal-fano} via the relation
\begin{equation}
 \phi = \arg{(q-i)} \;. \label{eq:mapping}
\end{equation}
Using this, the Fano formula [Eq.~(2) in the main text] can be rewritten to Eq.~(1). A similar mapping was found in Ref.~\cite{universal-fano}, where absorption lines of auto-ionizing helium have been studied. However, it is different by a factor 2 from the relation given in Eq.~(\ref{eq:mapping}). The reason for this is that, in contrast to Ref.~\cite{universal-fano}, our continuum is not the free space vacuum, but a cavity. The cavity vacuum also undergoes a phase shift upon a change of the incidence angle, and thus the relative phase is reduced, as we will show in the following.

We consider the phases of the different channels in more detail. To this end we specialize to the case $\Gamma\approx\gamma_\textrm{SR}\gg\gamma$ such that $q = \kappa/\Delta_C \in {\rm I\!R}$ and $\kappa = 2\kappa_R$. From Eqs.~(\ref{supp_eqn:R_C}) and (\ref{supp_eqn:R_N}) we obtain
\begin{eqnarray}
 R_C &=&  \frac{-i}{q+i}
      = \frac{-i}{\sqrt{1+q^2}} \, e^{i\phi} \;, \label{eq:phase_RC}\\
 R_N &=& -i \; \frac{1}{\epsilon+i}\, \frac{q-i}{q+i} 
      = -i \;  \frac{1}{\epsilon+i} \, e^{2i\phi} \label{eq:phase_RN}\;.
\end{eqnarray}
The phase of the cavity reflection changes as $\phi$ in dependence of $q$. Furthermore, from Eq.~(\ref{eq:phase_RN}) we find that the phase shift is $2\phi$ for the nuclear amplitude. Therefore, the relative phase between the continuum $R_C$ and the bound state $R_N$ is $(2-1) \phi = \phi$. To interpret the cavity as an interferometer, we attribute this relative phase to the cavity amplitude, such that the phase of the nuclear contribution to the reflectance depends only on the energy $\epsilon$. Neglecting a global phase, we find
\begin{equation}
 R = R_C + R_N  \propto r_C e^{-i \phi} + r_N e^{i\phi_N} \;,
\end{equation}
with
\begin{eqnarray}
 r_C &=& \frac{1}{\sqrt{1+q^2}} \;, \\
 r_N &=& \frac{1}{\sqrt{1+\epsilon^2}} \;, \\
 \phi_N &=& \arg{\left(\frac{1}{\epsilon+i}\right)} \;.
\end{eqnarray}

\section{Relation of the phase of off-diagonal nuclear density matrix elements with the phase of the detected light}
The nuclear phase $\phi_N$, which we determined in our experiment, can be identified with the phase $\phi_\rho$ of the off-diagonal elements of the two-level system (TLS) density matrix. This can already be seen by noting that the nuclear contribution to the reflected field amplitude $R_N \propto r_N \exp{(i \phi_N)}$ is directly proportional to off-diagonal density matrix element $ \rho_{eg} = |\rho_{eg}|\exp{(i \phi_\rho)} $ as shown in Ref.~\cite{qomodel}.

Here, we briefly outline this accordance using a different method. We found in the previous section  that the reflected field amplitude can be written as $r_C \exp{(-i \phi)} + r_N \exp{(i\phi_N)}$, such that an interference term $\propto \cos{(\phi+\phi_N)}$ is obtained in the intensity. For simplicity, we neglect the cavity environment, such that $\phi=0$, and describe the field amplitude by an alternate first principle approach. To this end, we consider the scattering of a plane wave field with positive frequency part
\begin{equation}
\Ein^{(+)} = E_{{\textrm{in}}} \vec{\epsilon}e^{i(\vec{k}\vec{r} - \omega t)}
\end{equation}
from a TLS. Here, $E_{{\textrm{in}}}$ is the field amplitude, $\vec{\epsilon}$ the polarization vector, $\vec{k}=k\hat{k}$ the wavevector, $\vec{r}=r\hat{r}$ the nucleus position, and $\omega$ the field frequency. The interaction with the TLS leads to a total field $\vec{E}^{(+)}  = \Ein^{(+)}  + \Escat^{(+)}$~\cite{E_out}
with 
\begin{equation}
\Escat^{(+)} = \frac{k^2}{4\pi\epsilon_0 r} e^{i(\vec{k}\vec{r}-\omega t )} \: \hat{S}_-\: (\hat{r}\times \vec{d})\times\vec{r}\,,\label{eq:s1}
\end{equation}
where $\vec{d}$ is the TLS dipole moment assumed parallel to the polarization, and $\hat{S}_-$ the transition operator from the upper to the lower state.
The total intensity registered by the detector is the given by $I\propto \langle \vec{E}^{(-)} \vec{E}^{(+)}\rangle$, and decomposes into a constant part $\xi = \langle \Ein^{(-)}\Ein^{(+)}\rangle +\langle \Escat^{(-)}\Escat^{(+)}\rangle$ and an interference part $\langle \Ein^{(-)}\Escat^{(+)}\rangle + \langle \Escat^{(-)}\Ein^{(+)}\rangle$.
Evaluated in forward direction, one finds
\begin{equation}
I\propto \xi + \frac{E_{\textrm{in}} \, d\, k^2}{4\pi \epsilon_0} \left ( \langle  \hat{S}_-\rangle + \langle  \hat{S}_+\rangle\right)
\end{equation}
with $\hat{S}_+ = \hat{S}_-^\dagger$. Using
\begin{equation}
\langle \hat{S}_-\rangle = \rho_{eg} = |\rho_{eg}| e^{i\phi_\rho}\,,
\end{equation}
where $\rho_{eg}$ is the off-diagonal TLS density matrix element, we obtain
\begin{equation}
I\propto \xi + \frac{E_{\textrm{in}} \, d\, k^2}{2\pi \epsilon_0} |\rho_{eg}| \: \cos(\phi_\rho)\,.
\label{eq:s2}
\end{equation}
As a result, we find that evaluated in forward direction, the scattered field $\Escat$ has a phase relative to the incident field given by the phase of $\rho_{eg}$, which using Eq.~(\ref{eq:s2}) can most easily be observed in the scattered light intensity in the interference term $\propto \cos(\phi_\rho)$. Comparing this result with the previously obtained interference term, we thus find the correspondence of the phases $\phi_N = \phi_\rho $.

\end{document}